\documentclass[twocolumn]{aastex631}
\usepackage{amsmath}
\usepackage{xspace}

\newcommand{\swift}{{\it Swift}\xspace}
\newcommand{\nicer}{\textit{NICER}\xspace}
\newcommand{\xmm}{{\it XMM-Newton}\xspace}

\newcommand{\cgs}{{\rm erg s$^{-1}$ cm$^{-2}$}\xspace}

\newcommand{\targetlong}{Swift J023017.0+283603\xspace}
\newcommand{\target}{Swift\,J0230+28\xspace}

\expandafter\ifx\csname natexlab\endcsname\relax\fi
\providecommand{\url}[1]{\href{#1}{#1}}
\providecommand{\dodoi}[1]{doi:~\href{http://doi.org/#1}{\nolinkurl{#1}}}
\providecommand{\doeprint}[1]{\href{http://ascl.net/#1}{\nolinkurl{http://ascl.net/#1}}}
\providecommand{\doarXiv}[1]{\href{https://arxiv.org/abs/#1}{\nolinkurl{https://arxiv.org/abs/#1}}}

\shorttitle{SwJ0230's eruptions have shutoff}
\shortauthors{Pasham et al.}

\graphicspath{{./}{figures/}}

\begin{document}

\title{Repeated Partial Tidal Disruptions and Quasi-Periodic Eruptions in SwJ023017.0+283603}
	\author[0000-0003-1386-7861]{Dheeraj Pasham}
	\affiliation{MIT Kavli Institute for Astrophysics and Space Research \\
		Cambridge, MA 02139, USA}

\author[0000-0003-3765-6401]{Eric R. Coughlin}
\affiliation{Department of Physics, Syracuse University, Syracuse, NY 13210, USA}

\author[0000-0002-2137-4146]{Chris~J.~Nixon}
\affiliation{School of Physics and Astronomy, Sir William Henry Bragg Building, Woodhouse Ln., University of Leeds, Leeds LS2 9JT, UK}

\author[0000-0001-6450-1187]{Michal Zaja\v{c}ek}
\affiliation{Department of Theoretical Physics and Astrophysics, Faculty of Science, Masaryk University,\\ Kotl\'a\v{r}sk\'a 2, 611 37 Brno, Czech Republic}

\author[0000-0002-4779-5635]{Petra Sukov\'a}
\affiliation{Astronomical Institute of the Czech Academy of Sciences, Bo\v{c}n\'{\i} II 1401, 141 00 Prague, Czech Republic}

\author[0000-0002-5760-0459]{Vladimir Karas}
\affiliation{Astronomical Institute of the Czech Academy of Sciences, Bo\v{c}n\'{\i} II 1401, 141 00 Prague, Czech Republic}

\author[0000-0002-0786-7307]{Thomas Wevers}
\affiliation{Space Telescope Science Institute, 3700 San Martin Drive, Baltimore, MD 21218, USA}


\author[0000-0002-6562-8654]{Francesco Tombesi}
\affiliation{Physics Department, Tor Vergata University of Rome, Via della Ricerca Scientifica 1, 00133 Rome, Italy}

\begin{abstract}
SwJ023017.0+283603 (SwJ0230) exhibited soft X-ray (0.3-1.0 keV) eruptions recurring roughly every 22 days. We present results from an extended monitoring campaign of SwJ0230 using Swift, NICER, and deep XMM-Newton observations. Our main findings are: 1) SwJ0230 did not display any eruptions during two 80-day periods (June-September 2023 and July-September 2024) of high-cadence monitoring with NICER and Swift, suggesting that the eruptions have ceased, implying an eruption lifetime of less than 536 days; 2)  quiescent/non-eruption emission is detected with XMM-Newton, with a 0.3-2.0 keV luminosity of 4$\times$10$^{40}$ erg/s (bolometric luminosity of $<$0.1\% Eddington assuming a black hole mass of 10$^{6-7}$ M$_{\odot}$), that is consistent with a thermal disk spectrum peaking at  0.11$^{+0.06}_{-0.03}$ keV; 3) SwJ0230 exhibited multiple, rapid eruptions (duration$<$5~hours, similar to quasi-periodic eruptions; QPEs), and there is tentative evidence that they recur, on average, on roughly the same timescale of 22 days. \target therefore exhibited (when active) both rapid, QPE-like outbursts and longer-duration outbursts, more akin to those from repeating partial Tidal Disruption Event (rpTDE)  candidates. These findings are difficult to explain with existing models that invoke an orbiter interacting with a persistent disk and those involving disk instabilities. We propose a hybrid model wherein an object of smaller mass (e.g., a Jupiter-sized planet) being repeatedly partially stripped and subsequently punching through its own, fallback-induced disk, can explain many of the observed properties, including the long-duration flares (from accretion), the short-duration outbursts (from the planet-disk interaction), and the turn-off of the flares (when the planet is totally stripped of gas). 
\end{abstract}



\section{Introduction}
Two classes of repeating extragalactic nuclear transients \citep[RENTs; see e.g.][for a review]{sukova2024} have been uncovered in the X-ray (0.2--10.0 keV) bandpass: those that repeat on ``long'' timescales of a few$\times$(months—years) \citep[e.g.,][]{wevers3, 2023A&A...669A..75L, 2024A&A...683L..13L} and those that have ``short'' recurrence timescales of a few$\times$(hours—days) \citep[e.g.,][]{gsndisc, rxdisc, qpe12, 2024SciA...10J8898P, 2024Natur.630..325P}\footnote{Within the context of this work the term short-period RENTs includes Quasi-Periodic Eruptions/QPEs and Quasi-Periodic Outflows/QPOuts \citep{2024SciA...10J8898P}.}. These discoveries were facilitated largely by all-sky optical and X-ray surveys, including ASAS-SN \citep{2014ApJ...788...48S}, Zwicky Transient Facility \citep{ZTF}, and eROSITA \citep{erosita}, that repeatedly scan large areas of the sky. While modulations are not seen in the optical/UV (except in the cases of ASASSN-14ko: \citealt{2021ApJ...910..125P, 2023ApJ...951..134P} and AT2018fyk/ASASSN-18ul: \citealt{2024ApJ...971L..31P}), optical surveys have been used to identify outbursts and provide the impetus for X-ray follow-up in several cases. 

A leading hypothesis for the origin of long-period RENTs is the repeated partial tidal disruption of a star (rpTDEs) \citep[e.g.,][]{2010MNRAS.409L..25Z, cufari22, wevers3, 2023A&A...669A..75L, 2024ApJ...971L..31P, evans23} (but see \citealt{wen24} for an alternative interpretation). 
For the short-period RENTs, two additional categories of models have been proposed: those that invoke an orbiting secondary object around a primary massive black hole (mass $\gtrsim$10$^{4}$ M$_{\odot}$)\citep[e.g., see ][]{2024SciA...10J8898P, 2024MNRAS.529.1440W, 2024ApJ...967...70R, 2023MNRAS.526L..31K} or disk instabilities \cite[e.g.,][]{marzenamodel, kaurmodel, panmodel, Raj:2021b}. In both classes of models, a newly formed accretion disk is necessary to explain the accompanying overall long-term decay \citep[e.g.,][]{aliveandbare, alivenkicking, qpe34, 2024ApJ...965...12C} (however,  \citealt{2024arXiv241100289P} where such a long-term decay is not evident for eRO-QPE2) or optical/X-ray outburst \citep[e.g.][]{2024SciA...10J8898P, guolo24, 2019qiz, bykov}, which can plausibly originate from a past tidal disruption of a star.


\targetlong (hereafter \target) was identified as an X-ray transient by the Living Swift-XRT Point Source catalog (LSXPS; \citealt{xrtlive}) and 
announced publicly on 22 June 2022 \citep{discovery}. The field of view was observed by \swift's X-Ray Telescope (XRT) on several occasions between 1 
December 2021 (MJD 59549) and 8 January 2022 (MJD 59587) to follow up a nearby supernova ($\sim$ 4$^{\prime}$ away from \target's position). X-ray emission was not detected from the nucleus of the host in any of these exposures, yielding a combined 0.3-2.0 X-ray upper limit of about 2$\times$10$^{-14}$ \cgs. Observations taken on 
22 June 2022 (MJD 59752) revealed an X-ray source with a flux of 8$\times$10$^{-13}$ \cgs, suggesting an enhanced flux by more than a factor of 40. The X-ray spectrum was soft and thermal with a temperature of roughly 0.12 keV. Based on the spatial coincidence of the source with the center of a galaxy at $z=0.036$ and a soft/thermal X-ray spectrum, it was initially reported to be a flare resulting from the tidal disruption of a star by a massive black hole \citep{discovery}. Daily XRT and \nicer monitoring of \target between 22 June 2022 (MJD 59752) and 1 February 2023 (MJD 59976) revealed the presence of intense X-ray eruptions, during which the observed 0.3-2 keV flux varied between 3$\times$10$^{-14}$ \cgs and $\approx$2$\times$10$^{-12}$ \cgs (see Fig.~3 of \citealt{evans23} and Fig.~1 of \citealt{guolo24}). The optical and UV emission remained roughly constant during this period (Fig.~1 of \citealt{guolo24})--consistent with the host galaxy.

In this work, we present an extended monitoring campaign of \target with \nicer and \swift, and a deeper \xmm exposure to characterize the quiescent emission. We discuss the data reduction/analysis in the appendix sections \ref{sec:swift}, \ref{sec:nicer} and \ref{sec:xmm}, and in section \ref{sec:res} we describe our main results. One of our main findings is that \target displayed no outbursts over a span of $>$160 days, during which multiple eruptions$\footnote{The words eruption, outburst and flare are used interchageably throught the text}$ should have been detected if the source was still active at the same flux level, suggesting that the eruptions have ceased and the mechanism responsible for powering them is no longer active. In Section \ref{sec:gasgiant} we provide a model, consisting of the repeated partial disruption of a low-mass object, that can explain every feature of this source, and we discuss the (generally disfavoring) implications of our results for various other models in Appendix \ref{sec:model-implications}. 

\begin{figure*}[ht]
    \includegraphics[width=\textwidth]{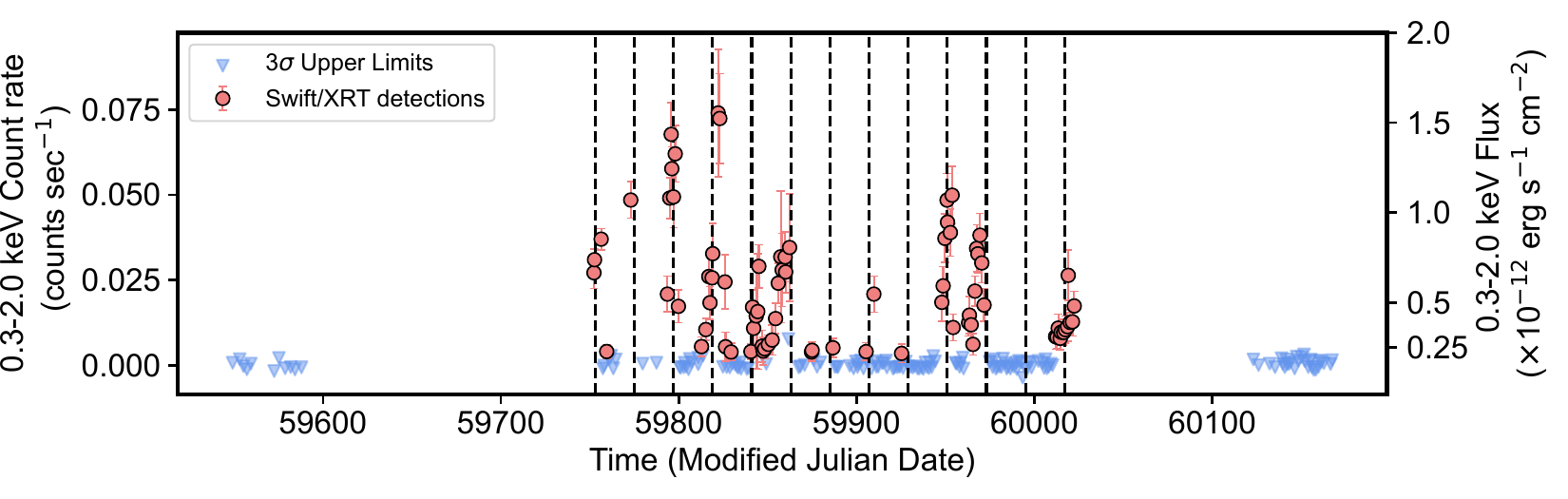}
    \caption{ \textbf{\target's unabsorbed 0.3-2.0 keV long-term light curve with \swift/XRT}. The data was binned on an obsID basis. The vertical dashed lines are uniformly separated by 22 days, the best-fit period estimated from the periodogram by \citet{guolo24}. }
    \label{Fig:fig1}
\end{figure*}

\begin{figure*}[ht]
    \includegraphics[width=\textwidth]{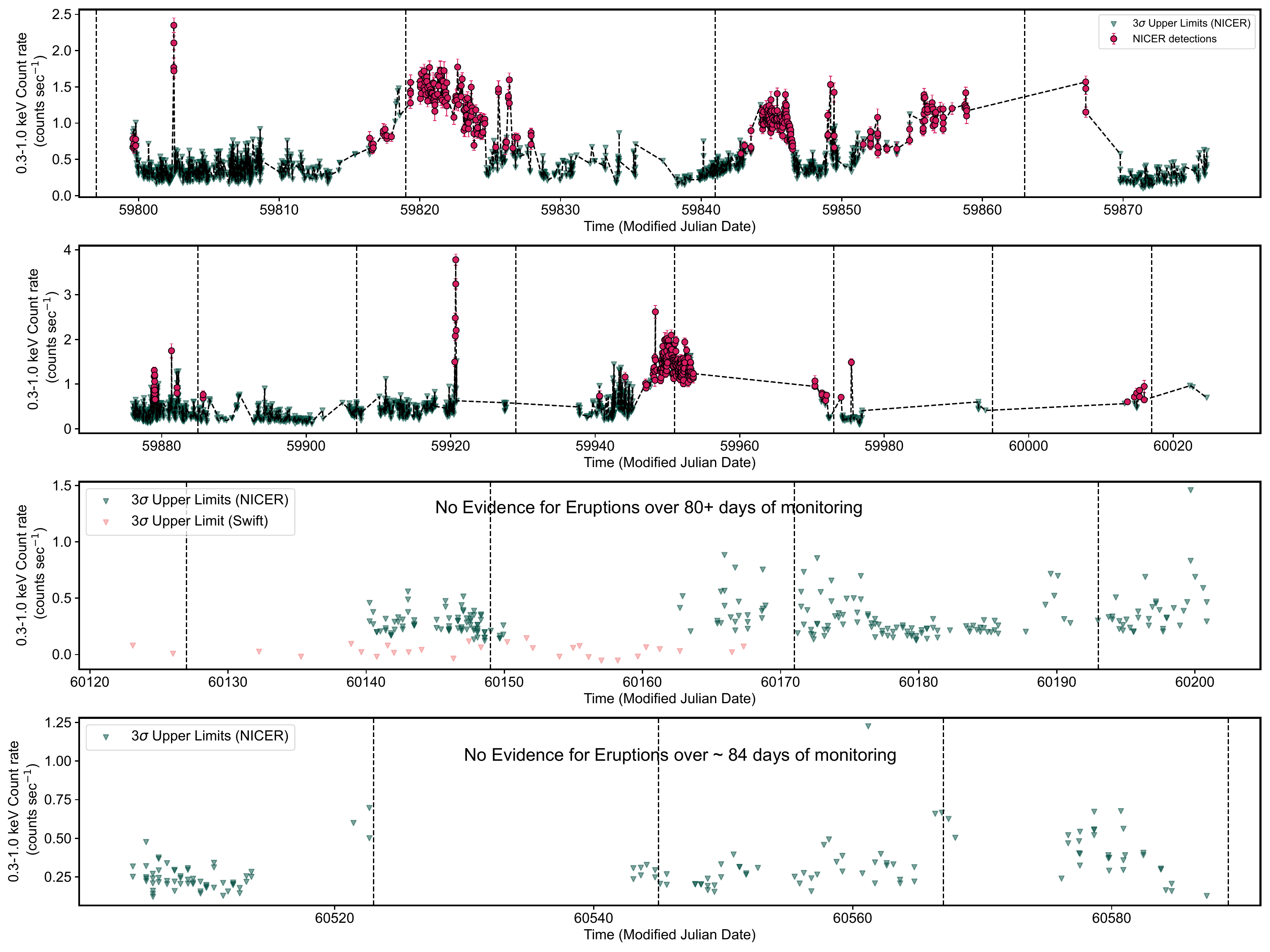}
    \caption{ \textbf{A summary of \target's long-term X-ray evolution}. The top two panels show the \nicer X-ray (0.3-1.0 keV) evolution while the bottom two panels show a period of \nicer and \swift with no evidence for eruptions. The vertical dashed lines are uniformly separated by 22 days with the same reference date as Fig.~\ref{Fig:fig1}. XRT's 3$\sigma$ upper limit is based on a scaling factor of 44 between 0.3-2.0 keV XRT count rate and 0.3-1.0 keV \nicer rate (See Appendix section~\ref{sec:nicer} for details on this estimate). Data prior to MJD 59976 was published in \cite{guolo24}.}
    \label{Fig:fig2}
\end{figure*}

\begin{figure*}[ht]
    \includegraphics[width=\textwidth]{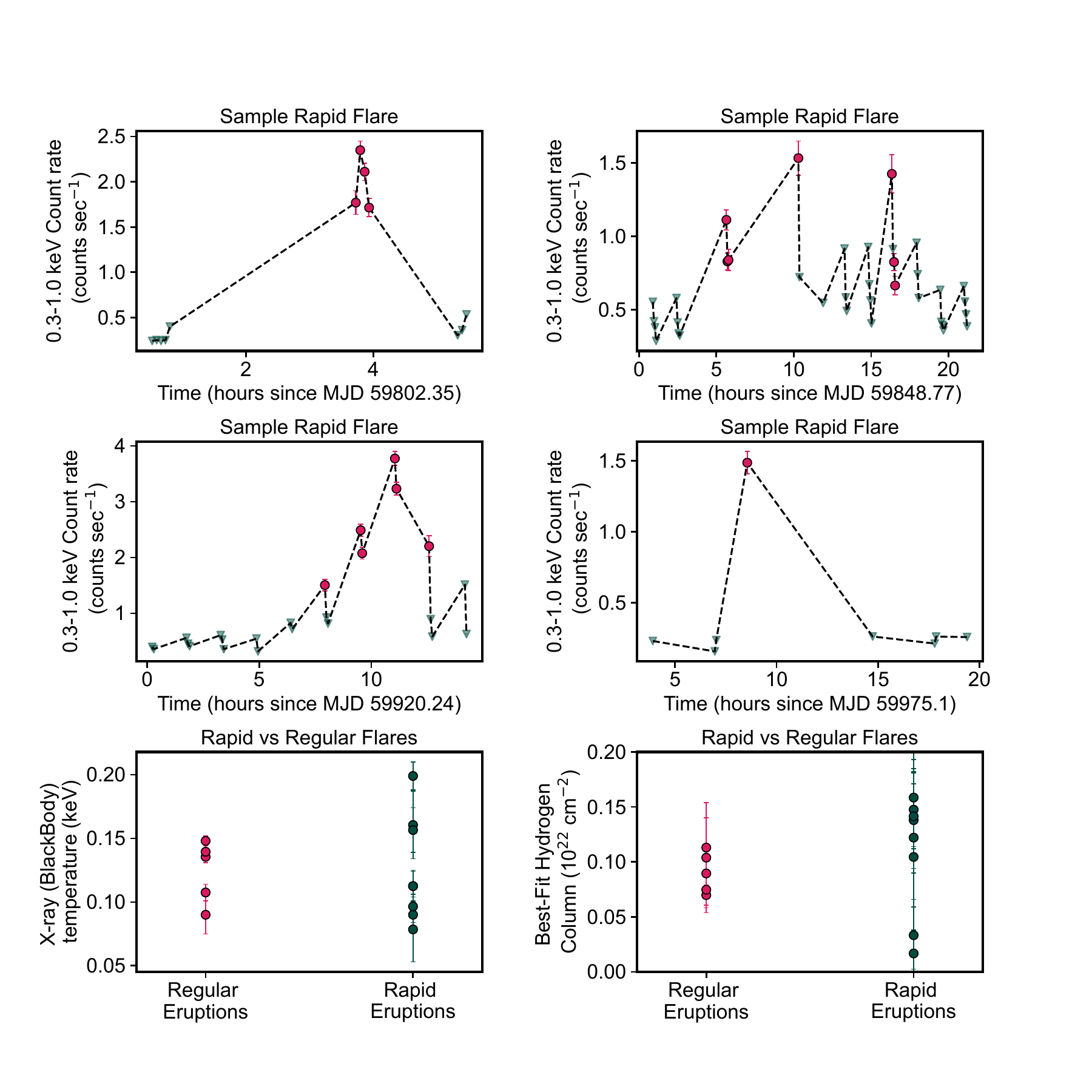}
    \caption{ \textbf{A summary of the rapid flares from \target}. The top four panels show sample rapid flares. {\it Bottom panels:} We compare the X-ray spectra of \target's rapid and regular flares. The y-axis of the left panel shows the best-fit temperatures near the peaks of the five well-sampled regular flares (red) and the nine rapid flares (green). The right panel  shows the best-fit neutral Hydrogen columns for the rapid (green) and regular (red) flares. One of the green data points is off the scale with a large errorbar (0.56$\pm$0.44) and is not shown in this plot. All spectra were fit with a single temperature blackbody, {\it ztbabs*zashift(bbody)} in {\it XSPEC}. All errorbars represent the 90\% uncertainty. }
    \label{Fig:fig3}
\end{figure*}

\section{Results}\label{sec:res}
We used data from three X-ray telescopes: \swift's X-Ray Telescope (XRT; \citealt{xrt}), \nicer \citep{nicer}, and \xmm's European Photon Imaging Camera (EPIC). A description of the data reduction and analysis procedures is given in the Appendix sections~\ref{sec:swift}, \ref{sec:nicer} and \ref{sec:xmm}. Here we discuss the main results.

\begin{figure*}[ht]
    \includegraphics[width=\textwidth]{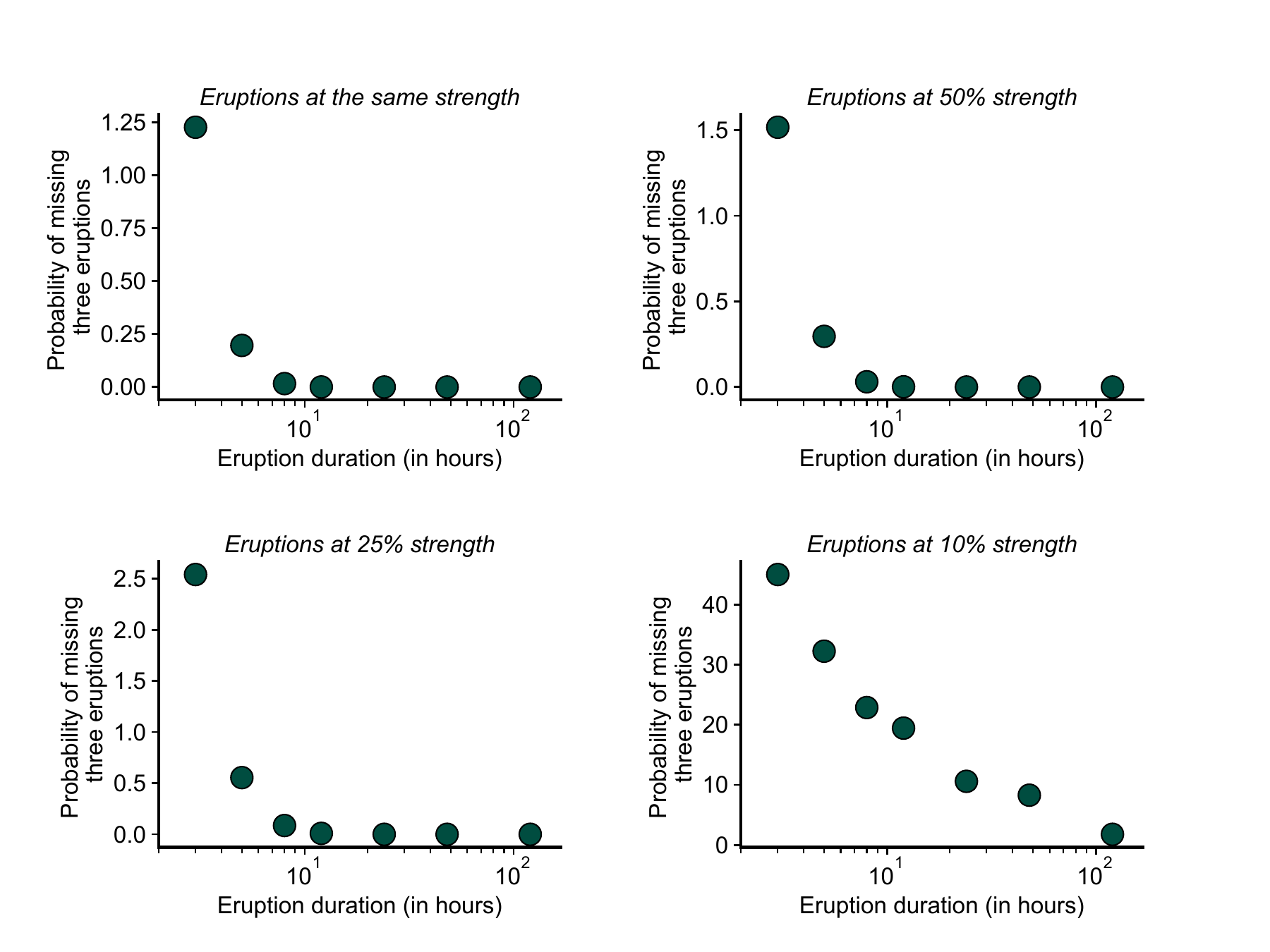}
    \caption{ \textbf{Results from simulations to assess the probability of missing three Gaussian shaped eruptions between MJDs 60140 and 60205}. The top-left panel shows the probability as a function of eruption duration assuming they had a peak flux comparable to the median of the peak fluxes during prior monitoring. The rest of the panels show the same plot but with different peak fluxes. The probability of missing regular (a few days long) eruptions is $<$10\% even if the eruptions have decreased in strength (peak flux) by an order of magnitude. The probability of missing rapid (a few hours long) flares is also low ($<$a few percent) if their peak fluxes were $>$1/4$^{\rm th}$ their previous strength.}
    \label{Fig:fig4}
\end{figure*}

\target's long-term 0.3-2.0 keV XRT light curve is shown in Fig.~\ref{Fig:fig1}. While eruptions are present between MJDs 59750 and 60025 \citep{evans23, guolo24}, no eruptions are evident in the data taken between MJDs 60123 and 60168.

Next we extracted \nicer's 0.3-1.0 keV background-subtracted light curve (Fig.~\ref{Fig:fig2}) which shows: 1) no eruptions between MJDs 60140 and 60205 and 2) the presence of rapid eruptions with typical duration of $\lesssim$5 hours (see top and middle panels of Fig.~\ref{Fig:fig3}).

\begin{figure*}[ht]
    \includegraphics[width=\textwidth]{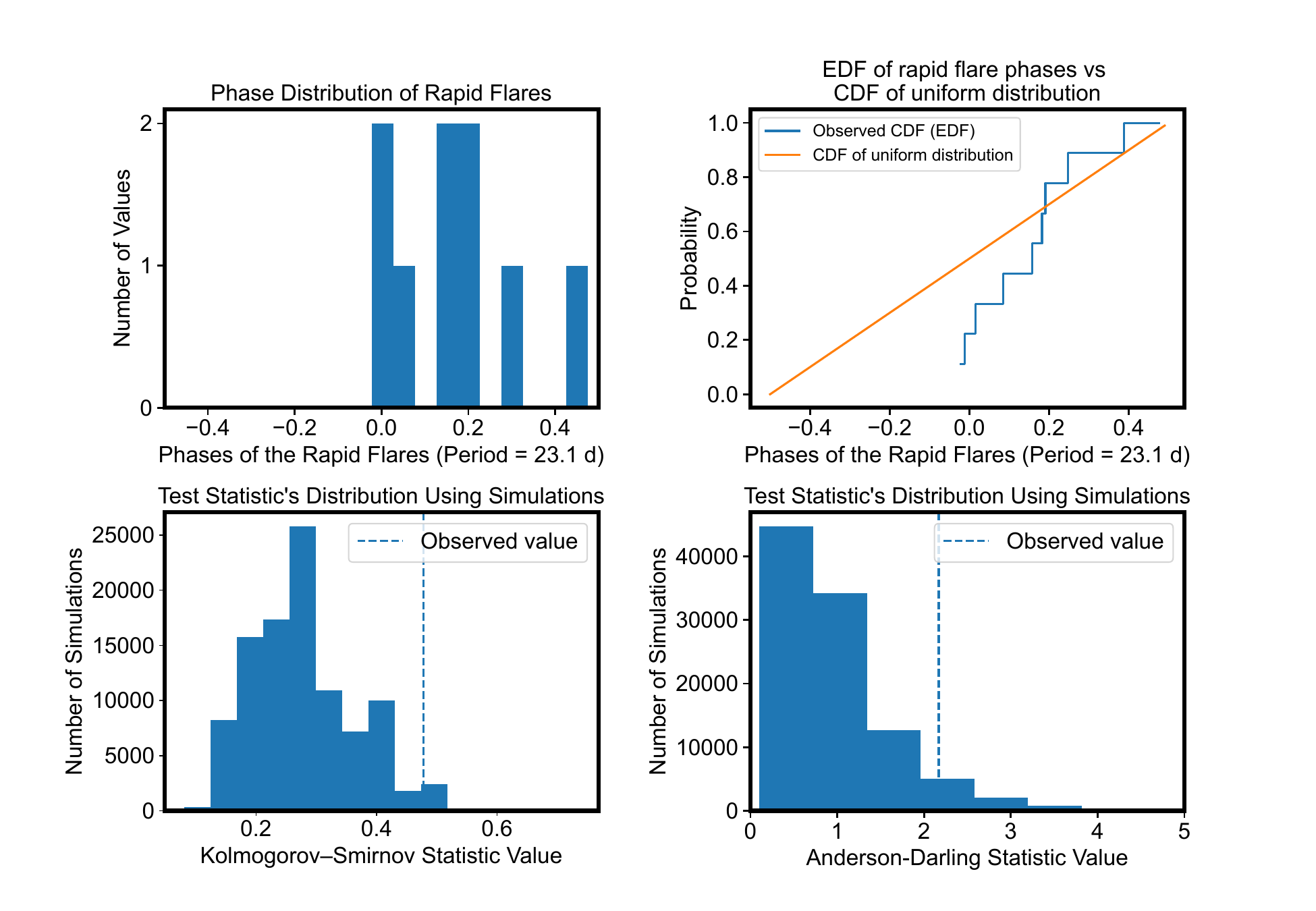}
    \caption{ \textbf{Summary of quantifying the quasi-periodicity of the rapid flares}. The top-left panel shows the distribution of the phases of the rapid flares: seven out of nine values are within a phase range of 0-0.2, i.e., 20\% deviation from strict periodicity. These values are consistent with known QPE recurrence periods. They are normalized to be between -0.5 and 0.5. The top-right panel shows the empirical distribution function (EDF) of the phases of the rapid flares (histogram) along with the cumulative distribution function (CDF) of a uniform distribution between -0.5 and 0.5. The bottom-left and the bottom-right panels shows the distributions of the Kolmogorov–Smirnov and the Anderson-Darling statistics against a uniform distribution, respectively. Only 2.5\% and 6\% of the K-S and Anderson-Darling statistic values are above the  observed values (dashed vertical lines), i.e., the chance probability of having 9 random rapid eruptions aligned to the level observed in the data is at most 6\%.}
    \label{Fig:rapidstat}
\end{figure*}

Combined with XRT data taken around the same time we conclude that the regular eruptions reported in \citet{evans23, guolo24} with a mean duration of a few days are likely not present in data past MJD 60140, or -- if they are present -- they have decreased in amplitude by more than a factor of 10 (see Fig.~\ref{Fig:fig4} and section~\ref{sec:prob} for further details). A low-activity phase similar to that exhibited between MJDs 59865 and 59945 (see Fig.~\ref{Fig:fig1}) is also unlikely, unless the rapid eruptions are less than a factor of 4 weaker (see Fig.~\ref{Fig:fig4} and section~\ref{sec:prob})

The eruptions began sometime between MJDs 59587 and 59752, and ceased between MJDs 60023 and 60123. As such, the phase during which eruptions occurred was shorter than 536 days -- one of the tightest constrains on the lifetime of such systems (see \citealt{aliveandbare,alivenkicking,2024ApJ...965...12C, qpe34} for similar constrains on QPE sources). 

The \nicer light curve also exhibits rapid flares whose times (in MJD) are: 59802.494$\pm$0.049, 59825.62$\pm$0.064, 59826.359$\pm$0.129, 59848.972$\pm$0.031, 59879.037$\pm$0.061, 59882.233$\pm$0.160, 59920.702$\pm$0.032, 59948.30$\pm$0.040 and 59975.505$\pm$0.080 days. The uncertainties (in days) are half the block sizes from the Bayesian blocks algorithm of \citet{blocks2}. The block sizes can be treated as a proxy for the duration of these rapid eruptions, whose median and standard deviation values are 3 hours and 2 hours, respectively. For comparison, the same algorithm gives a median (standard deviation) of 6.1 (1.4) days for six of the best-sampled eruptions with XRT. These numbers are comparable to \citet{guolo24}'s estimate using asymmetric Gaussian modeling. 

Combining with the rapid flare detected in XRT data around MJD 59909.71, the median time between consecutive flares is 23.1 days--consistent with the recurrence time of the regular eruptions. We assess the statistical significance of the quasi-periodicity of the rapid flares in Appendix section~\ref{sec:statsig} and are summarized in Fig.~\ref{Fig:rapidstat}. The chance probability of clustering at any period is small ($<$6\%). Based on this, we proceed with the assumption that the rapid flares are also quasi-periodic.

Next, we extracted the spectra of nine of the best-sampled rapid flares, and they are consistent with thermal emission at temperatures between 0.05-0.21 keV. These are similar to the temperatures at the peaks of the regular broad eruptions \citep{guolo24}, and their best-fit neutral column densities are also comparable (see bottom panels of Fig.~\ref{Fig:fig3}).

Finally, we analyzed the \xmm data and clearly detect a point source coincident with the position of \target (see Fig.~\ref{Fig:xmm}). We extract its spectrum and measure the logarithm of its unabsorbed 0.3-2.0 keV flux of -13.9$\pm$0.1, from which we calculate the amplitude of the eruptions to be 191$\pm$66 (see also Appendix section~\ref{sec:xmm}). For comparison, the \swift/XRT quiescent values reported by \citep{evans23} were (1-3)$\times$10$^{-14}$ \cgs during active eruptions, suggesting that the quiescent level may have declined by a factor of a few similar to a few known QPEs \citep{2024ApJ...965...12C, alivenkicking}.



\subsection{Assessing the chance probability of missing activity due to sampling}\label{sec:prob}
The mean duration of the regular and the rapid eruptions prior to MJD 60100 was 3 days (see Fig.~5 of \citealt{guolo24}) and 3 hours, respectively. Given the observing cadence and the upper limits in the second from bottom panel of Fig.~\ref{Fig:fig2}, we compute the probability of missing eruptions with a wide range of duration (3 hours to 5 days) and peak fluxes (Fig.~\ref{Fig:fig4}). To compute these numbers we place a Gaussian eruption of a certain width and peak flux at 5000 randomly chosen points between MJD 60140 and 60205. We chose this time period as it provided the densest monitoring of \target without a detection. Then we estimate the number of times this Gaussian intersects the observing windows and is also above the upper limits. This can then be turned into the chance probability of missing an eruption of a certain peak flux and width. Then, this number is cubed to compute the probability of missing three such eruptions expected over a $\sim$65 day time span. We perform this exercise for four peak fluxes for the Gaussian: 1) median of the observed peak fluxes of eruptions of 1.4 cps, 2) 50\%, 3) 25\% and 4) 10\% of the median of the observed peak fluxes. The results are shown in Fig.~\ref{Fig:fig4} and should be considered upper limits, as we do not include the probability of a non-detection in the more recent observations between MJDs 60503 and 60587. 

During prior monitoring there was a time window of $\sim$ 80 days between MJDs 59865 and 59945 where \target showed weak eruption activity:  a few rapid eruptions around MJDs 59880, 59910 and 59921 (see Fig.~\ref{Fig:fig1} and \ref{Fig:fig2}). From the above simulations, it can be seen that the probability of missing 3 such rapid flares, assuming the same strength as before, 50\% and 25\% of previous strength, is $<$1.5\%, $\approx$1.5\%, and $\approx$2.5\%, respectively. This number increases to 40\% if the eruptions have decreased in strength to 1/10$^{\rm th}$ of the original eruptions. We conclude that the eruptions have likely turned off or decreased in magnitude by more than a factor of 10, with only a few percent likelihood that we may be missing them due to monitoring gaps. 

\subsection{Comparison to other RENTs}
Three QPE sources (GSN069, eRO-QPE1 and eRO-QPE3) \citep{alivenkicking, aliveandbare, qpe34} and two of the best-sampled repeating TDEs exhibit a declining trend in the peak outburst flux with time \citep{2024ApJ...971L..31P, 2024A&A...683L..13L}. While \target's eruptions have disappeared, it is difficult to definitively say if there was a gradual decline in eruption strength, especially because of the lack of data between its last major outburst and the complete cessation of emission (see Fig.~\ref{Fig:fig1}). However, \target's average amplitude (ratio of peak to the quiescent flux) is higher than that of QPEs (see Fig. 5 of \citealt{guolo24}) but is below the two X-ray rpTDEs AT2018fyk/ASASSN-18ul \citep{2024ApJ...971L..31P} and eRASSt J045650.3{\ensuremath{-}}203750 \citep{2023A&A...669A..75L}. Based on the amplitude alone, it is unclear if \target can connect QPEs and rpTDEs, though it lies firmly between the two classes in peak luminosity (as also noted in \citealt{evans23, guolo24} based on the recurrence timescale of the flares).

On the other hand, \target exhibits both QPE-like outbursts that are $\sim$ a few hours in duration, and longer-duration flares that are $\sim$ few days in duration that are reminiscent of those from rpTDEs. Therefore, instead of being a ``bridge'' between QPEs and rpTDEs, \target may instead possess physical characteristics that facilitate the emergence of \emph{both} types of eruptions. The model we propose in Section \ref{sec:gasgiant} below does precisely this.

\section{discussion}\label{sec:dis}
Our main observational findings are: 1) the absence of eruptions in new monitoring data, which constrains \target's eruption-phase lifetime to less than 536 days, 2) the confirmation of several rapid flares that last only for a few hours, i.e., shorter than 1/10$^{th}$ of the regular eruptions reported earlier in \citep{evans23, guolo24}, and are consistent with a repetition timescale of 22 days and 3) the detection of quiescent flux, which allows us to measure the eruption amplitude to 191$\pm$66. In addition, there is a time window of about 80 days (between MJDs 59865 and 59945) where \target only exhibited rapid eruptions (see Fig.~\ref{Fig:fig1}, \ref{Fig:fig2} and ~\ref{Fig:fig3}). Any potential model will need to explain all these properties.

RENT models can be put into two broad categories: 1) inner-disk instabilities, and 2) repeated encounters between a secondary object and an accretion disk.

\subsection{Implications for disk instability models}\label{sec:instab}
There are several processes that may operate within a disk to produce quasi-periodic electromagnetic flares: 1) radiation pressure instability \citep[e.g.,][]{marzenamodel}, 2) inner disk (rigid-body) precession \citep[e.g.,][]{2012PhRvL.108f1302S}, 3) disk tearing \citep{nixon2012}, and 4) shock front oscillations \citep{2017MNRAS.472.4327S}. The first two models were already discussed and  disfavored by \citet{guolo24} because the underlying disk is accreting at $<$0.1\% of the Eddington limit (see Appendix \ref{sec:xmm}; the radiation pressure instability is thought to arise at high accretion rates \citep{marzenamodel}. Lense-Thirring precession is disfavored based on the eruption profiles and the observed irregularity \citep{guolo24}. We discuss disk tearing and shock front oscillations in Appendix sec.~\ref{sec:model-implications}, and disfavor them owing to their inability to explain the rapid flares. 

\subsection{Implications for models invoking repeated disk collisions by an orbiter in near circular orbit}
Models where eruptions arise from shocks produced by an object punching through a cold/optically thick accretion disk \citep{alessiamodel,linial23,Yao:2024rtl} are also unlikely, given the low bolometric luminosity of $<$ 0.1\% of the Eddington limit where the disk is expected to be Advection Dominated Accretion Flow (ADAF)-like (geometrically thick and optically thin) \citep[e.g.,][]{1994ApJ...428L..13N,1996ApJ...471..762A}. Even if the disk is thin at this accretion rate, the interactions of the star with the disk leading to expanding shocks would yield a luminosity that is too small for typical parameters \citep{linial23},
\begin{align}
  L_{\rm erupt}&\sim 1.2 \times 10^{40} \left(\frac{R_{\star}}{1\,R_{\odot}} \right)^{2/3} \left(\frac{M_{\bullet}}{10^{6.6}\,M_{\odot}} \right)\times \,\notag\\
  &\times \left(\frac{\dot{m}}{10^{-3}} \right)^{1/3} \left(\frac{P_{\rm erupt}}{22\,\text{days}} \right)^{-2/3}\,{\rm erg\,s^{-1}}\,,
  \label{eq_lum_erupt}
\end{align}
where $R_{\star}$ is the stellar radius and $P_{\rm erupt}$ is the mean recurrence timescale. $\dot{m}$ and M$_{\bullet}$ are the accretion rate and black hole mass, respectively. This value can be boosted by increasing the ellipticity of the orbit but the main issue is to produce two types of eruptions (see next section).

Distinctly, \citet{2021ApJ...917...43S}'s orbiting model involves an ADAF--type accretion flow around an SMBH that is repeatedly ``pushed'' around by an orbiter. However, it is again difficult to envision the presence of both types of  flares in this setup.


\citet{itaimodel1} calculated a single-Extreme Mass Ration Inspiral (EMRI) rate of $10^{-5} - 10^{-7}$ yr$^{-1}$, such that it is difficult (probabilistically) to generate two such objects on near-circular orbits simultaneously. 
We nevertheless consider the scenario of two co-orbiting stars around an SMBH \citep{2022ApJ...926..101M,guolo24} and how such a system could potentially explain both the longer regular flares (when the two stars meet) and the rapid flares (when the stars hit the disk during each short orbit). For more details, see Appendix~\ref{sec:rep-encoun}. 


\subsection{The repeated partial disruption of a gas giant in a highly eccentric orbit}
\label{sec:gasgiant}
\target displays both long-duration ($\gtrsim$ days) outbursts that are reminiscent of repeating partial TDEs \citep{payne21, wevers3, liu23a, somalwar23} as well as short-duration ($\sim$ hours) outbursts that are more similar to QPEs. If we attribute the more extended outbursts to accretion events onto the black hole, the overall energetics are consistent with a small amount of mass ($\lesssim 10^{-5}M_{\odot}$) removed from the object (for accretion efficiencies of $\sim 10\%$). There are also noticeable ``gaps'' in the frequency of the longer outbursts, where the 22-day repetition cycle would predict an outburst that was actually absent. Here we consider a physical model that can plausibly explain these features, being the repeated partial disruption of a low-mass object\footnote{A massive object would require a fine-tuned pericenter distance to remove a small amount of its mass (see \citealt{alivenkicking} for an analogous argument that disfavors a white dwarf). One would then need to invoke a third body (i.e., Kozai-like oscillations) on an orbit of only marginally greater semimajor axis to stop the eruptions on short timescales, because the star would be in the classic tidal dissipation regime (tides would remove orbital energy at the expense of tidally heating the star; \citealt{cufari23}), and the gravitational binding energy of the orbit is significantly greater than the binding energy of the object itself, i.e., one would completely destroy the object prior to ejecting it \citep{cufari23}.} (putatively a gas giant) that impacts its own accretion disk to produce the QPE-like outbursts.

The fallback time of debris in a TDE is usually the one that determines the rate at which material is accreted onto (or at least supplied to) the black hole, and scales as the dynamical time of the partially disrupted object, where the dynamical time is $\tau_{\rm dyn} \simeq 1/\sqrt{G\rho}$ with $\rho$ the average density. Since gas giants and main sequence stars have comparable densities, their dynamical times are also comparable. The proportionality constant is the square root of the mass ratio, such that if $m$ is the mass of the partially disrupted object, the fallback time is
\begin{equation}
    T_{\rm fb} \simeq \tau_{\rm dyn}\sqrt{\frac{M_{\bullet}}{m}},
\end{equation}
with $M_{\bullet} \simeq 10^6 M_{\odot}$ the SMBH mass. With $\tau_{\rm dyn} \simeq 0.5 \textrm{ hr} \simeq 0.02$ days, this gives $T_{\rm fb} \simeq 20$ days for a sun-like star, whereas a Jupiter-mass object -- with a mass of $\sim 0.001 M_{\odot}$ and radius $\sim 0.1 R_{\odot}$ -- yields a fallback time of $\sim 2$ years. 

It would appear that this timescale is too long to be consistent with the $\sim 20$-day recurrence time of the \target. However, the large discrepancy between the fallback time and the orbital time $T_{\rm orb}$, presumed to be on the order of $T_{\rm orb} \simeq 20$ days, implies that it is actually the latter, not the former, that sets the mass supply timescale. In particular, this system is in the regime where the specific binding energy to the SMBH significantly outweighs the spread in energy imparted to the debris from the tidal field of the SMBH, where the latter is
\begin{equation}
    \Delta\epsilon = \frac{Gm}{R}\left(\frac{M_{\bullet}}{m}\right)^{1/3},
\end{equation}
with $R$ the radius of the object. The binding energy of the orbit $\epsilon$, on the other hand, is simply obtained from the energy-period relation of a Keplerian orbit. For numbers appropriate to a Jupiter-like object and a $10^6 M_{\odot}$ black hole, the spread in the specific binding energy is $\Delta\epsilon \simeq 2\times 10^{-5} c^2$, while $\epsilon \simeq 3\times 10^{-4} c^2$, i.e., $\Delta\epsilon/\epsilon \simeq 0.1$. In this limit, the total duration of each outburst is set by the difference in the orbital times of the most- and least-bound debris, which is (to leading order in the ratio $\Delta\epsilon/\epsilon$)
\begin{equation}
    \Delta T \simeq T_{\rm orb}\frac{3\Delta\epsilon}{\epsilon} \simeq 6\textrm{ days}.
\end{equation}
The period of $T_{\rm orb} = 20$ days can be obtained via Hills capture of a sufficiently hardened binary, the hardened nature required for the binary to survive in the galactic nucleus (see \citealt{cufari22} and \citealt{evans23} in the context of this specific event). 

The expected outburst duration of $\sim 6$ days is in good agreement with the observations of \target. In this model, each outburst consists of the accretion of two tails of debris, one more bound and the other less bound, with the planet returning in between. In other rpTDE sources the return of the core was interpreted to correlate with the prompt shutoff of accretion \citep{wevers3}, owing to the fact that for very weakly bound orbits, the region of the debris stream excised of mass by the gravitational field of the core grows with time as $\propto t^{2/3}$ \citep{coughlin19}. This approximation should be upheld until roughly the apocenter is reached, after which this region grows more rapidly owing to the difference in acceleration of the least-bound fluid element within the debris stream and the core itself. Therefore, even if the mass ratio is very small, this region excised of mass (akin to the Hill sphere until apocenter is reached) can grow to an appreciable size over timescales of many months to years. Here, however, the mass ratio is even smaller (than for a star and a black hole) and the orbital time is considerably shorter than other rpTDEs, and hence it is unlikely that this region will be large enough to significantly modify the mass supply rate.

If the radiative efficiency of accretion is $1-10\%$, the accreted mass per large outburst is $\sim 1-10\%$ the mass of a Jupiter-sized object. It therefore seems plausible that a Jupiter-mass object could lose up to tens of percent of its total mass per pericenter passage, thus resolving the fine-tuning problem that would arise if the object were solar-like, i.e., it would need to be situated at a very precise radius to lose a miniscule fraction of its mass on each encounter. 

This configuration could also explain the discontinuation of eruptions as well as the overall cessation of emission: recent simulations by \citet{bandopadhyay24b} have shown that certain stars are capable of withstanding many, repeated tidal encounters with a supermassive black hole, losing only a small fraction of their mass during each pericenter passage and the amount of mass stripped (per encounter) declining with time. The origin of this stellar survivability is that the energy imparted via tides is deposited primarily in the outer layers of the star (where the amplitude of the tidal field is largest), and those outer layers are subsequently stripped by the black hole, while the high-density core of the star remains relatively unperturbed and provides the self-gravitational field needed to withstand the black hole's tidal shear. Gas giants are thought to consist of a gaseous envelope alongside an inner, rocky core, where the latter could contain up to ten percent of the total planet mass (e.g., \citealt{militzer08}). This core could act analogously to that of a star if its tensile strength is sufficiently high, meaning that as the outer layers of the gas giant are stripped, the core is effectively unaltered. Furthermore, if a substantial fraction of the envelope is stripped during the first few encounters, the planet's average density could initially increase as more gas is concentrated near the rocky core, resulting in a lower effective $\beta$ (equal to the tidal radius divided by the pericenter distance). It would then require a number of additional encounters, during which little to no mass would be lost, to tidally heat and inflate the remaining envelope until it were susceptible to additional tidal stripping. The accretion outbursts terminate when most-to-all of the mass is removed from the planet, leaving the rocky core behind as an orbiter. 

The QPE-like outbursts could originate from the interaction between the orbiter and the disk, similar to the models that employ this interpretation for QPE-only systems \citep{linial23,franchini23}. The long-short cycles exhibited by QPEs have been suggested to arise from the two interactions of the (in that case) star per orbit, there being a slight difference in time from one outburst to the next owing to the (implied) mild eccentricity of the orbit. Here, however, the orbit of the perturber is highly eccentric: with a pericenter distance equal to $\sim$ the tidal radius of $\sim 50 GM_{\bullet}/c^2 \simeq 7.4\times 10^{12}\textrm{ cm}$ (for an object with solar-like density) and a semimajor axis of $\sim 2.3\times 10^{14}$ cm (implied by an orbital period of 20 days and a black hole mass of $10^6M_{\odot}$), the orbital eccentricity is $e \simeq 0.97$. Instead of two outbursts per orbit, the orbiter could intersect the disc only once near pericenter if the pericenter vector of the planet's orbit (i.e., the vector pointing from the black hole to the pericenter of the orbiter) is nearly within the plane of the accretion disk, with a correspondingly weak intersection near apocenter (assuming the disc has had time to viscously spread that far). Alternatively, if the pericenter vector is nearly orthogonal to the plane of the disc, we would expect two outbursts in close succession of one another, separated in time by roughly half the orbital time at pericenter. With a pericenter distance on the order of tens of gravitational radii and a rapidly spinning black hole, the accretion flow could rapidly align to the black hole spin, while the orbiter precesses both nodally and apsidally by some amount per orbit. It could therefore be the case that the short-amplitude outbursts are variable in time throughout the encounter, as the degree of apsidal precession (which is on the order of tens of degrees for fairly relativistic pericenters) determines the (mis-)alignment with the disc plane, but we would expect their recurrence time to -- on average -- reflect the orbital period of the orbiter and the time between larger outbursts. 

The speed of the orbiter near pericenter is $\sim few\times 0.1$ c (note that this is much larger than for near-circular orbits with the same period), which should be approximately the same speed as the shock formed as it impacts the disk. With a radiation-pressure dominated shock and a temperature of $0.2$ keV, this implies a disk density of $2\times 10^{-9}$ g cm$^{-3}$, which is consistent with the densities inferred from other rpTDEs (see \citealt{wevers3}). Additionally, with a luminosity of $\sim 2\times 10^{42}$ erg s$^{-1}$ and a temperature of $0.2$ keV, the radius of the emitting surface is $\sim 9.8\times 10^{9}$ cm, which is comparable to that of a gas giant (Jupiter's radius is $\sim 7\times 10^{9}$ cm). Finally, if the orbiter interacts with $10\%$ of its mass per orbit, or $\sim 10^{-4} M_{\odot}$, and is moving at $\sim 0.2 c$, then the kinetic energy of the outflowing material as the orbiter impacts the disc -- assuming the shock speed is comparable to the orbital speed, which should be a good approximation for the large disparities in density -- is $E_{\rm kin} \simeq 0.5\times 10^{-4}M_{\odot}\times\left(0.2 c\right)^2 \simeq 3.5\times 10^{48}$ erg, which corresponds to $\sim 2\times 10^{44}$ erg s$^{-1}$ if all of the kinetic energy could be radiated over the $\sim 5$-hour duration of the short bursts (note that, again because of the highly eccentric orbit and high speed near pericenter, this is larger than the luminosities estimated for near-circular orbits; cf.~Equation \ref{eq_lum_erupt}). This implies that the radiative efficiency of the shock-heated gas is $\sim few\%$, which is consistent with observations of core-collapse supernovae \citep{2019NatAs...3..697I}. 

Finally, a natural question to ask is why other rpTDEs do not show QPE-like outbursts, and this could be related to the difference between the accretion timescale and the orbital timescale: rpTDEs have so far been discovered around very massive black holes ($\gtrsim$10$^{7}$ M$_{\odot}$; ASASSN-14ko: \citealt{payne21}; AT2018fyk: \citealt{wevers3}; eRASStJ045650.3-203750: \citealt{liu23}), such that the pericenter distance is highly relativistic, and their recurrence times ($\sim$ orbital times) are on timescales of months-years. In such systems, one would expect a relatively short viscous time relative to the orbital time (indeed, this is invoked to explain the rapid shutoffs observed in, e.g., AT2018fyk; \citealt{wevers3}), implying that the disc is rapidly drained and is no longer present as the star returns to pericenter. QPE systems that harbor a persistent, underlying accretion flow with which an orbiter interacts could be in the other extreme limit, where the long viscous time allows many, repeated interactions with the orbiter. \target could be a relatively rare, in between case, such that the orbital pericenter is close enough to the black hole to permit relatively rapid accretion -- and thus power the longer-duration flares -- but the viscous time is sufficiently long to permit the presence of a reservoir of gas near pericenter with which the orbiter interacts, i.e., the viscous time near pericenter is comparable to the orbital time.


\section{Summary}
We presented results from an extended monitoring of \target and report three new aspects: 1) discovery of the quiescent emission, which allows us to constrain the Eddington ratio of the underlying disk to be $<$0.1\% of the Eddington values after accounting for the uncertainties in black hole mass from M-$\sigma$ of 6.6$\pm$0.40 and bolometric color-correction; 2) the cessation of the eruptions, suggesting that they possess a lifetime of less than 536 days in \target; and 3) several epochs of rapid flares that repeat roughly on the same timescale as the regular eruptions. We propose that these rapid eruptions are akin to QPEs and the regular eruptions are similar to repeating partial TDEs, i.e., \target exhibits both types of repeating signals. We propose a model involving the repeated partial disruption of a Jupiter-sized object, which can produce the longer-duration eruptions from the accretion of tidally stripped material, and the rapid eruptions from the remnant core punching through its own fallback disk. Other systems, including GSN069, AT2019vcb and AT2019qiz, exhibit both tidal-disruption-like flares and QPEs (respecitvely \citealt{alivenkicking, bykov, 2019qiz}); the distinction with this source is that it is the same object that is providing both types of outburst -- in the latter two (and in the models proposed by \citealt{linial23,franchini23}) the disk and longer-duration flare is supplied by a third body, namely a passing and tidally stripped star.

\newpage

\section*{Acknowledgments}
D.R.P was funded by a NASA/XMM grant for this work (award number 035212-00001). E.R.C.~acknowledge support from NASA through the Astrophysics Theory Program, grant 80NSSC24K0897. C.J.N.~acknowledges support from the Science and Technology Facilities Council (grant No. ST/Y000544/1) and from the Leverhulme Trust (grant No. RPG-2021-380). MZ acknowledges the GA\v{C}R JUNIOR STAR grant No. GM24-10599M for financial support.
VK acknowledges the OPUS-LAP/GA\v{C}R-LA bilateral grant (2021/43/I/ST9/01352/OPUS 22 and GF23-04053L).


\appendix
\setcounter{table}{0}
\renewcommand{\thetable}{A\arabic{table}}
\setcounter{figure}{0}
\renewcommand{\thefigure}{A\arabic{figure}}

\section*{Data}\label{sec:dataapp}
Within the context of this work, we use a standard $\Lambda$CDM cosmology with parameters H$_{0}$ = 67.4 km s$^{-1}$ Mpc$^{-1}$, $\Omega_{\rm m}$ = 0.315 and $\Omega_{\rm \Lambda}$ = 1 - $\Omega_{\rm m}$ = 0.685 \citep{2020A&A...641A...6P}. Using the Cosmology calculator of \cite{2006PASP..118.1711W}, \target's redshift of 0.036 corresponds to a luminosity distance of 165.5 Mpcs.

The data reduction was performed using the standard HEASoft version 6.33.2 and the latest calibration files were used for \swift X-Ray Telescope (XRT), \nicer and \xmm. 

\section{\swift/XRT Data Reduction and Analysis}\label{sec:swift}
\swift observed the target on 239 occasions between 1 December 2021 and 11 August 2023. These obsIDs were downloaded from publicly-accessible HEASARC archive: \url{https://heasarc.gsfc.nasa.gov/cgi-bin/W3Browse/w3browse.pl}. Two of the observations did not have any Photon Counting data and was excluded from further analysis. The rest 237 obsIDs were reduced using the standard data reduction procedure using the {\it xrtpipeline} as outlined on \swift data analysis threads: \url{https://www.swift.ac.uk/analysis/xrt/xrtpipeline.php}. After that we used the {\tt xrtlccorr} tool to estimate a mean 0.3-2.0 keV exposure and background-corrected count rate per obsID (see  Fig.~\ref{Fig:fig1}).

To convert from 0.3-2.0 keV background-subtracted count rate to intrinsic 0.3-2.0 keV flux we extract an average spectrum using all obsIDs with count rates great than 0.003 cps. We binned the spectrum using the {\it ftgrouppha} tool with {\it grouptype=optmin} and {\it groupscale=25}. We then fit this spectrum in {\it XSPEC} \citep{xspec} using a blackbody and a powerlaw component modified by Galactic absorption: {\it tbabs*zashift*cflux*(bbody+pow)} in {\it XSPEC} which resulted in a best-fit $\chi^{2}$/degrees of freedom (dof) of 12.7/12. The Hydrogen column of {\it tbabs} component was fixed to a value of 0.074$\times$10$^{22}$ cm$^{-2}$ \citep{nHMap}. The best-fit bbody temperature was 0.113$\pm$0.005 keV while the powerlaw index was 2.5$^{1.0}_{-1.8}$. The logarithm of the average unabsorbed 0.3-2.0 keV flux in units of \cgs is -11.98$\pm$0.03 which corresponds to a count rate of (1.90$\pm$0.05)$\times$10$^{-2}$ cps. Using this conversion ratio we estimated the flux during individual obsIDs. 



\section{\nicer Data Reduction and Analysis}\label{sec:nicer}
As of 15 October 2024, the HEASARC has 248 \nicer obsIDs. The actual number is slightly higher but those other obsIDs do not yield any exposure with standard data filters (see below). There is an ongoing monitoring as part of an approved cycle 6 \nicer GO program (PI: Guolo; Program ID 7131). Here we include all data available on the HEASARC as of 15 October 2024. We started analysis by downloading the data from the HEASARC and reduced it using the standard procedures outlined on \nicer data analysis guide: \url{https://heasarc.gsfc.nasa.gov/docs/nicer/analysis_threads/}, i.e., {\tt nicerl2} followed by {\tt nicerl3-spect}. To extract background-subtracted light curves we used the {\tt nicerl3-lc} tool with a binsize of 250 s following the standard steps outlined here: \url{https://heasarc.gsfc.nasa.gov/docs/nicer/analysis_threads/nicerl3-lc/}.



To quantitatively assess the \nicer light curve we apply the Bayesian Blocks algorithm \citep{blocks2} to the entire \nicer light curve. In addition to the broad (several days long) eruptions that appear roughly every 22 days, the algorithm identified nine instances of rapid flares which lasted less than five hours. 

To assess these flares we extracted their spectra and modeled them using the SCORPEON background in {\it XSPEC}. They all are consistent with a thermal blackbody with a median (standard deviation) temperature of 0.11 (0.04) keV. Typically, \nicer background flares are ``hard" and almost never thermal. There was also an instance of such a flare with lower-cadence \swift/XRT \citep{evans23, guolo24} (See Fig.~\ref{Fig:fig1}). Considering these we conclude that these are real.

There were four instances around MJD 60143, 60190,  60199, and 60550 where {\tt nicerl3-lc} suggests a marginal detection. We inspected them manually and found that these were instances where the entire excess above the background was in a single spectral bin between 0.5-0.6 keV. This is from a well-known blend of foreground Oxygen and Ne features from Earth's atmosphere. Allowing the Oxygen normalization in the SCORPEON background modeling gets rid of this excess above the background. Consequently, we conclude that these three marginal detections are not related to the source. 

Given \nicer's high-cadence data there were multiple GTIs within 0.1 days of XRT observations. We compared the background-subtracted 0.3-1.0 keV \nicer count rates with 0.3-2.0 keV XRT count rates in those epochs, and estimated a scaling factor of 44 between the two quantities. 

\section{\xmm Data Reduction and Analysis}\label{sec:xmm}
\xmm observed \target for 49.8 ks on 25 January 2024 (MJD 60334) as part of an approved guest observer program to measure its quiescent spectrum (PI: Pasham). We reduced this obsID 0923950101 using the standard \xmm reduction procedure, i.e., with the {\it epproc} and the {\it emproc} XMMSAS tools. Background flaring was present and was removed by assessing the light curve of a nearby background region free of any point sources. This reduced the total exposure to roughly 33 ks. 

\begin{figure*}
    \centering
    \includegraphics[width=\textwidth]{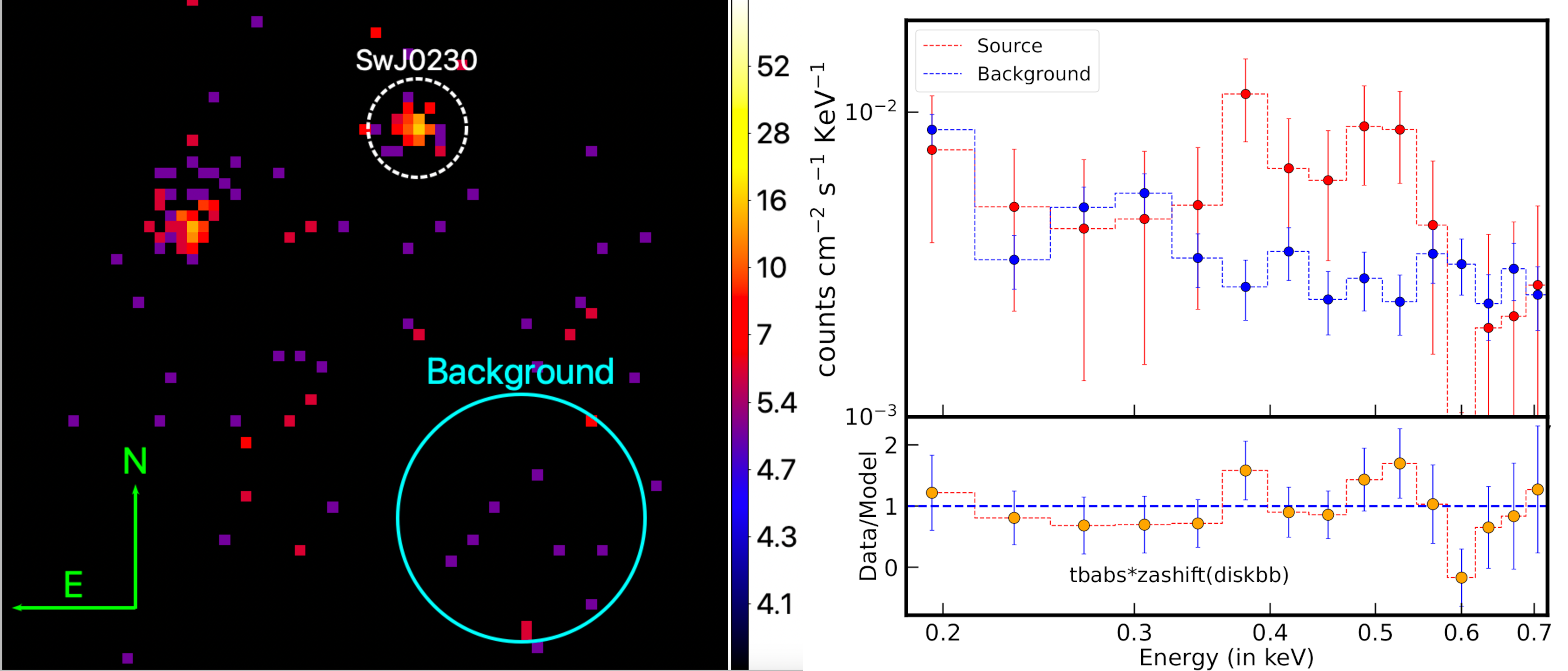}
    \caption{\textbf{Summary of the \xmm data analysis.} The left panel shows an EPIC (pn+MOS) 0.2-1.0 keV image of \target's field of view after the eruptions turned off. The dashed white circle is centered on \target's position and has a radius of 25$^{\prime\prime}$. The green arrows pointing North and East are each 50$^{\prime\prime}$ in length. {\it Right:} The top panel shows the source and background spectra while the bottom panel shows the ratio of the data to the best-fit thermal disk model. }
    \label{Fig:xmm}
\end{figure*}

A source at the position of \target was clearly evident in the raw 0.2-1.0 keV image (left panel of Fig.~\ref{Fig:xmm}). We extracted a pn spectrum following the procedure outlined on \xmm data analysis webpage: \url{https://www.cosmos.esa.int/web/xmm-newton/sas-threads}. Source events were extracted from a circular region with a radius of 25$^{\prime\prime}$ while the background was estimated using a nearby 50$^{\prime\prime}$ radius circular region. Because the background starts to dominate beyond 0.7 keV we only used the 0.2-0.7 keV bandpass and modeled the spectrum with a thermal disk: {\it tbabs*zashift*cflux*diskbb} in {\it XSPEC}. This resulted in a best-fit C-stat/dof of 10/11 with a disk temperature and the logarithm of the inferred unabsorbed 0.3-2.0 keV flux of 0.11$^{+0.05}_{-0.03}$ keV and -13.9$\pm$0.1 \cgs, respectively. The background-subtracted 0.2-0.7  keV count rate was (2.5$\pm$0.4)$\times$10$^{-3}$ cps. 

We estimate a median (standard deviation) value of the 0.3-2.0 keV unabsorbed peak flux of eruptions using four of the best-sampled \nicer eruptions to be 2.4(0.6)$\times$10$^{-12}$ \cgs. Due to \nicer's large effective area this has a smaller uncertainty but is consistent with the measurement from XRT data of (1.8$\pm$0.9)$\times$10$^{-12}$ \cgs.

\subsection{Eddington Ratio during quiescence}
The unabsorbed 0.3-2.0  keV inferred luminosity is 40.58$\pm$0.14. Integrating the best-fit disk over 1 ev to 10 keV gives a value of 41.07$^{+0.36}_{-0.28}$. Using a logarithm of the central supermassive black hole mass of 6.6$\pm$0.40 derived from host galaxy stellar velocity dispersion \citep{guolo24} implies that the  Eddington ratio during quiescence is (2.1$\pm$0.5)$\times$10$^{-4}$. The 90\% uncertainty value was estimated using the \texttt{bootstrap} functionality in {\it Scipy}: \url{https://docs.scipy.org/doc/scipy/reference/generated/scipy.stats.bootstrap.html}. This includes the errorbars on the integrated 1 eV to 10 keV luminosity and the 0.4 dex uncertainty in the black hole mass. If we are extra-conservative and include an additional color-correction factor of 5, the bolometric Eddington ratio is still below 0.1\%.

\section{Assessing the Statistical Significance of the quasi-periodicity of the rapid eruptions}\label{sec:statsig}
To test if there is any regularity in the occurrence of the rapid flares we first computed the time difference between consecutive flares and obtained a modulo with respect to the median of the time between consecutive rapid flares. This gives a phase with respect to the median time between flares of 23.1 d, and is shown in the top left panel of Fig.~\ref{Fig:rapidstat}. For reference, if the flares were strictly periodic the distribution would have a single bin at 0 and if they are quasi-periodic they would cluster near 0 with a distribution whose width is a measure of period jitter. Fig.~\ref{Fig:rapidstat} (top left) shows that 8 out of the 9 phase values are clustered in a narrow phase range of 0 and 0.3 suggesting a potential quasi-periodicity. 

Next, we devised a scheme to assess the probability that randomly placed flares would produce the same level of clustering. We start by comparing the empirical distribution function (EDF) of the observed phases with the cumulative distribution function (CDF) of a uniform distribution between -0.5 and 0.5 (Fig.~\ref{Fig:rapidstat}, top-right). Qualitatively, they do not appear to align.

To quantify the discrepancy between the CDF of a uniform distribution and the EDF of observed data we devise the following scheme:
\begin{itemize}
    \item We randomly chose 9 times between MJDs 59799 and 60020 and compute their phase distribution as done for the real rapid flares above.
    \item Then we extract the EDF of these simulated phases from the above step and compute the Kolmogorov–Smirnov (K-S) statistic against a uniform distribution. Conceptually, this statistic measures the maximum deviation between the two distributions.
    \item We repeat the above two steps 100,000 times and get a distribution of the K-S statistic between a uniform distribution and simulated random rapid flares.
    \item Finally, we compute the K-S statistic between the observed rapid flares and a uniform distribution. The distribution from above step is plotted along with this value in the bottom left panel of Fig.~\ref{Fig:rapidstat}.
    \item We also repeat the above steps using the Anderson-Darling statistic which is shown in the bottom-right panel of Fig.~\ref{Fig:rapidstat}.
\end{itemize}

The percentage of points in the K-S and Anderson-Darling based approaches that are above the observed value are 2.5\% and 6\%, respectively. Based on the above analysis, the likelihood that 9 random flares are separated as they do in the real data is less than 6\%.

\section{Implications for other models}
\label{sec:model-implications}

\subsection{Disk tearing}
Disk tearing is driven by an instability that occurs in strongly warped disks \citep{nixon2012,dogan2018,raj2021a,drewes2021}. The instability causes the warp to sharpen and the surface density to decrease sharply, breaking the disk into discrete planes. When the disk is subject to a forced precession, such as the Lense--Thirring effect from a spinning black hole or the gravitational torque from a companion star, the broken parts of the disk precess effectively independently. Any subsequent radial spreading of the precessing rings leads to collisions of gas orbits that shock heat the gas and rob it of rotational support causing inflow and enhanced accretion at smaller radii. The geometry of the system, and particularly the line of sight to the observer, also play a significant role in determining the observational appearance of the disk \citep{Raj:2021b}. The fundamental timescale on which variable emission occurs is the precession timescale, with the precession responsible for inducing the disk warp. For disks around black holes, precession via the Lense-Thirring effect requires the disk to be misaligned to the black hole’s spin.

The black hole mass in \target is expected to be around $10^{6.5} M_\odot$ based on host galaxy scaling relations, but due to the low host galaxy velocity dispersion there is significant uncertainty in this value. There is no constraint on the black hole spin. The two types of flares in \target are broadly categorized as short outbursts (lasting a few hours) and seen every $\sim$22 days and long outbursts (which emit almost all of the energy) lasting a few days and repeating less regularly but on a similar recurrence time. We therefore consider the timescales that might be expected to result from disk tearing in this system.

The Lense-Thirring precession timescale, for a spin of order unity and a black hole mass of $10^{6.5} M_\odot$ is $t_{\rm p} \sim 2.5(R/R_{\rm g})^3\,{\rm s}$ where $R_{\rm g} = GM/c^2$ is the gravitational radius of the black hole. To match with the $\sim 22$ day period in the lightcurve, the implied radius that is undergoing Lense-Thirring precession is around $100 R_{\rm g}$, which is within the region that is expected to be susceptible to the disk tearing instability for typical parameters of disks around supermassive black holes \citep{nixon2012}. Using their estimate of where the disk might break, with a spin of order unity and a disk viscosity parameter $\alpha \approx 0.1$ \citep{martin19}, then disk tearing would require $H/R \lesssim 0.01$. While AGN disks are typically much thinner than this, SwJ0230 is not a strong AGN \citep{evans23}. They measure the quiescent $0.3-2$\,keV luminosity to be $\approx 10^{41}$\,erg/s, which implies an Eddington ratio of $\sim 0.001(k/M_6)$ where $k$ is the bolometric correction factor from the $0.3-2$\,keV luminosity and $M_6$ is the black hole mass in units of $10^6M_\odot$. If the black hole mass is low or the bolometric correction significant then the disk may be thin enough for disk tearing to occur. 

The 22-day timescale may therefore be explained by tearing of the disk. However, we would expect the flare duration and the recurrence timescale in a disk tearing model to be similar as they are both driven by precession. The two timescales need not be exactly the same, but a duration that is significantly shorter than the recurrence period -- here a factor of at least 100 shorter -- is not expected. Thus we rule out disk tearing as being responsible for the short duration flares. The longer duration flares, in which most of the energy is emitted, are more in line with the expected emission from disk tearing \citep[e.g.][]{Raj:2021b}. It is worth noting that the long duration flares do not follow a strictly period pattern, and may therefore be more naturally explained by an instability than a periodic orbiting object (but see the discussion in \citealt{evans23}). The cessation of the flares would require a change in disk structure that renders the disk stable, which may result from a drop in the accretion rate and a transition to a thick, radiatively inefficient flow.

In summary, disk tearing cannot explain the short duration bursts, but may be able to explain the long duration bursts if the black hole mass is on the low end and there is a bolometric correction factor from the X-ray flux of at least a few. Disk tearing may result from the disk being misaligned to the spin of the black hole. The short duration flares could then be produced by a companion object, perhaps responsible for producing the disk, orbiting through the tearing disk; this scenario would generate a model that is an extension of that proposed by \cite{franchini23} with the disk warping and/or tearing rather than rigidly precessing.

\subsection{Shock front oscillations as flaring mechanism}
The possibility of the origin of eRO-QPE1 flares coming from oscillating shocks in low angular momentum flow was discussed in \citet{2024ApJ...963L..47P}. Here, the period is more than a factor of 20 longer than in the case of eRO-QPE1, where the period varied between 0.8 -- 1.0 days during five epochs. Taking into account the mass estimate $M_{\bullet}\in(10^{6.2},10^{7})M_\odot$, the observed variability corresponds to frequency $f \in (4 \times 10^{-6},3 \times 10^{-5}){c^3}/{(GM_{\bullet})}$. As shown in Fig.~5 and Tab.~2 and 3 of \citet{2017MNRAS.472.4327S}, such values correspond to the shock front location of several tens to few hundreds of gravitational radii (50 - 200 $GM_{\bullet}/c^2$) with large amplitude of the oscillations of the shock front position. The repeated large inflation and deflation of the shock bubble can explain the observed high amplitude of flares.

The spectrum of such flow would correspond to the hot flow model studied in detail in \citet{2014ARA&A..52..529Y}. The measured flux in 0.3-2.0 keV in quiescence corresponds roughly to $\log(\nu L_\nu) = 5\times10^{40}$  erg s$^{-1}$. Comparing with the model hot flow spectra shown in Fig.~1 of \citet{2014ARA&A..52..529Y}, such value at energy $\sim$ 1 keV could correspond to hot flow around SMBH with the estimated mass $\log(M_\bullet)=6.6\pm0.4$ with the Eddington ratio at the level $\dot{m} \sim 10^{-4}-10^{-3}$, which is in agreement with the estimate given in Section~\ref{sec:xmm}. For such low accretion rate, the spectrum of hot flow can consist of three distinct peaks corresponding to synchrotron radiation, inverse Compton component and bremsstrahlung, where the inverse Compton component may fit into the observed band.

However, during the flares, the measured flux in 0.3-2.0 keV yields values around $\log(\nu L_\nu) = 1\times10^{43}$  erg s$^{-1}$. Suppose this value is due to an accretion event. In that case, the accretion rate increases up to several per cent of Eddington accretion rate, which is hardly achievable in the hot accretion flow regime, as can be seen from Fig.~1 of \citet{2014ARA&A..52..529Y}. Moreover, for higher values of accretion rate, the individual peaks in the spectrum are smeared, hence it would be more challenging to explain the spectral shape. However, the shock changes the solution downstream, yielding higher density and temperature with lower inward velocity, which could help to explain this difference. A detailed study of the dissipative processes at the shock front in low angular momentum flows is needed to confirm or disprove the ability of this model to explain the flaring state of \target, which we leave to future work. Nevertheless, two types of eruptions are difficult to envision in this model.

\subsection{Implications for models that invoke repeated mass transfer from secondary objects}\label{sec:rep-encoun}

As discussed in \citet{guolo24}, a potential  mechanism for triggering the eruptions is the Roche-lobe overflow model of an evolved star on a mildly eccentric orbit \citep{krolikmodel}. Assuming the energy dissipation close to the innermost stable circular orbit, the mass loss per eruption can be estimated using,
\begin{align}
    \Delta M_{\star}&\simeq \frac{12 D P_{\rm erupt}L_{\rm erupt}}{c^2}\,\notag \\
    &\sim 9.6 \times 10^{-6} \left(\frac{D}{0.25}\right) \left(\frac{P_{\rm erupt}}{22\,\text{days}}\right) \left(\frac{L_{\rm erup}}{3\times 10^{42}\,{\rm erg\,s^{-1}}}\right)\,M_{\odot}\,,  
    \label{eq_mass_loss}
\end{align}
where we adopted $D=0.25$ and $L_{\rm erupt}=3\times 10^{42}\,{\rm erg\,s^{-1}}$ for the eruption duty cycle and the peak X-ray luminosity, respectively. The stellar mass-loss rate estimated for the duration of one eruption is approximately $\dot{M}_{\star}\sim 12L_{\rm erupt}/c^2\sim 6.3 \times 10^{-4}\,M_{\odot}{\rm yr^{-1}}$, which is comparable to the mass-loss rate via stellar winds for massive stars, such as Wolf-Rayet stars. 

For a single star on a mildly eccentric orbit with the orbital period of 22 days, the stellar radius is set by the tidal (Hill) radius, $R_{\star}\simeq 23\,R_{\odot} (P_{\rm orb}/22\,\text{days})^{2/3} (m_{\star}/1\,M_{\odot})^{1/3}$, which implies that the Roche-lobe overflowing star should be evolved. The soft X-ray eruption emission can then be generated via oblique shocks in stream-stream collisions as discussed by \citet{krolikmodel} and \citet{guolo24}. However, the main problem with the scenario involving a single Roche-lobe overflowing star is the efficient removal of the angular momentum from  the matter inspiralling from a few 100 $r_{\rm g}$.

\begin{figure}
    \centering
    \includegraphics[width=0.8\textwidth]{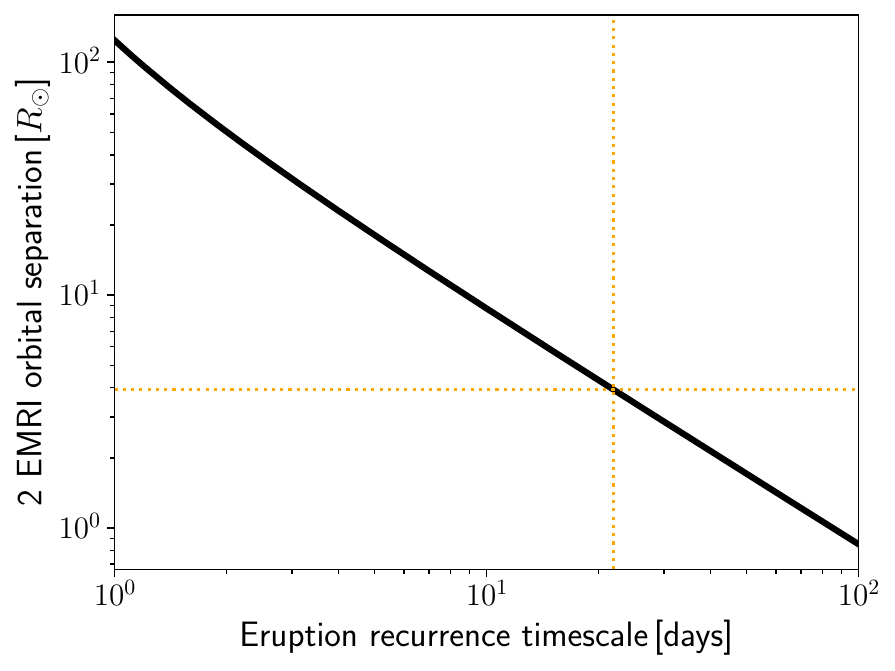}
    \caption{Semi-major axis difference (in Solar radii) for the double-EMRI model as a function of the X-ray flare recurrence timescale (in days). We fixed $R_{\star}=1R_{\odot}$, $m_{\star}=1\,M_{\odot}$, and $M_{\bullet}=10^{6.6}\,M_{\odot}$. Dotted orange lines represent the recurrence timescale and the suitable component separation for Swift J0230+28.}
    \label{fig_2EMRI}
\end{figure}

The problem concerning efficient angular-momentum removal is partially mitigated in the scenario involving two stars with slightly different masses where the more massive one has inspiralled faster due to larger gravitational-wave losses \citep{2022ApJ...926..101M}. The recurrence timescale of the eruptions is set by the pattern speed instead of the orbital timescale, or in other words, the synodic period when the two stars approach each other. In this scenario, the flares can be powered via the enhanced Roche-lobe overflow from the more massive star (see the description above) or via clumps produced in wind-wind collisions \citep[though the clump masses are typically lower than required by Eq.~\eqref{eq_mass_loss}, see e.g.][]{2020MNRAS.493..447C}. In comparison with the original set-up discussed in \citet{2022ApJ...926..101M}, \citet{guolo24} proposed the model involving two co-orbiting stars on tightly bound orbits with semi-major axes $a_1\simeq a_2$, with $a_2>a_1$ and a small difference of the order of a few stellar radii, $\Delta a=a_2-a_1=fR_{\odot}$. Since at least one of the stars is Roche-lobe overflowing, the semi-major axes are set by the tidal radius, $a_1/r_{\rm g}\simeq 40.8 (R_{\star}/1\,R_{\odot}) (m_{\star}/1\,M_{\odot})^{-1/3}(M_{\bullet}/10^{6.6}\,M_{\odot})^{-2/3}$. 

In this set-up, the quiescent X-ray emission is given by the overflowing stellar material inspiraling towards the SMBH, while the regular  eruptions reflect the enhanced overflow due to the second star that is periodically approaching the first one. The thermal X-ray emission could be produced in stream-stream oblique shocks \citep{krolikmodel} or potentially from a compact accretion disk formed from the overflowing material \citep{2022ApJ...926..101M}, and the regular eruptions could be triggered by circularization shocks when an enhanced matter inflow hits the disk \citep{2024arXiv240714578Y}. For the two co-orbiting Solar-type stars scenario, the synodic period for the orbital periods of $P_1$ and $P_2=P_1+\Delta t$ is $P_{\rm s}=P_1 P_2/(P_2-P_1)=(P_1^2+P_1 \Delta t)/\Delta t\simeq 22$ days, i.e., it is equal to the eruption recurrence timescale. For Swift J0230+28, we can infer that $\Delta a=3.92\,R_{\odot}$ (see also Fig.~\ref{fig_2EMRI}). This implies the orbital periods of $P_1\simeq 0.3705$ days and $P_2\simeq 0.3768$ days for the two stars (both $\sim$9 hours), yielding the fly-by periodicity of $P_{\rm s}\simeq 22$ days consistent with the regular flare periodicity. 

This setup can also address the presence of the two types of flares -- regular ones, which are induced during the time of closest approach of the two stars, repeating every 22 days, and rapid eruptions induced by the fast-orbiting star with an orbital period of $\sim 9$ hours as it interacts with its own disk. While this models predicts that rapid eruptions should recur roughly every 9 hours, they should be enhanced only once every 22 days.   

For the system of two co-orbiting stars, there is a natural mechanism to stop the flaring. In case they are misaligned with respect to the SMBH equatorial plane, they are subject to Lense-Thirring nodal precession, which results in the misalignment timescale of \citep{2022ApJ...926..101M},
  \begin{equation}
        \tau_{\rm prec}=607\,\left(\frac{a_{\bullet}\sin{\iota}}{0.004}\right)^{-1} \left(\frac{M_{\bullet}}{10^{6.6}\,M_{\odot}} \right)^{-2} \left(\frac{a}{40.8\,r_{\rm g}} \right)^3\text{days}\,.
    \end{equation}
 That means for the model involving two interacting stars (EMRIs) the flaring duration in Swift J0230+28 constrains the SMBH spin to lower values of $a_{\bullet}\lesssim 0.01$ (a dimensionless SMBH spin and adopting $\iota=25^{\circ}$ for the stellar orbital inclination).

\newpage
\bibliographystyle{aasjournal}
\bibliography{ms}

\end{document}